\documentclass[a4paper,10pt]{article}
\usepackage{graphicx}
\begin{document}

\title{\bf Analysis and Predictions of Social Phenomena via social media using Social Physics method}
\author{Akira Ishii\\
\sl \small Department of Applied Mathematics and Physics, Tottori University\\
Tottori 680-8552, Japan\\
\small ishii@damp.tottori-u.ac.jp}
\date{}
\maketitle

\begin{abstract}
As a method of analyzing and predicting social phenomena using social media as data, we propose a method using a mathematical model of hit phenomenon which is the theory of social physics. I could explain the transition of the number of social media written in movies, TV dramas, Music Concerts, Pokemon GO and proposed a method that can be used for analysis and prediction of social phenomena. 
\end{abstract}


\section{Introduction}

\noindent In the present age where consumer behavior remains on record through the Internet, purchase records and action records for huge quantities of consumers are left. When analyzing it, there are many cases where it is possible to incorporate natural science methodology such as physics, apart from conventional social science. In this paper, we propose a method based on social physics for analyzing and forecasting social phenomena, and possibly applying it to marketing etc. by using the voices of society 's people recorded by blogs and Twitter as data. Social physics is a new frontier of physics alongside economic physics, but if there is a huge amount of data, the methodology of physics that has been the subject of experimental data on natural phenomena can also be applied to social science. Therefore, in recent years, social physics is a field that is progressing rapidly.　　

Let's see how the number of daily postings on Twitter written about Japanese famous tennis player Nishikori Kei as an example of how the increase and decrease in the number of social media writes shows an increase or decrease in people's interests in society . As shown in the figure, writing on Kei Nishikori written on Twitter is increasing in conjunction with his tennis game day. Therefore, from this figure his game of tennis will know when. As described above, the number of writes to social media reflects the movement of society, so analysis of society becomes possible from analysis of the number of writes.

\begin{figure}[!h]
\centering
\includegraphics[height=6cm, bb=0 0 720 540]{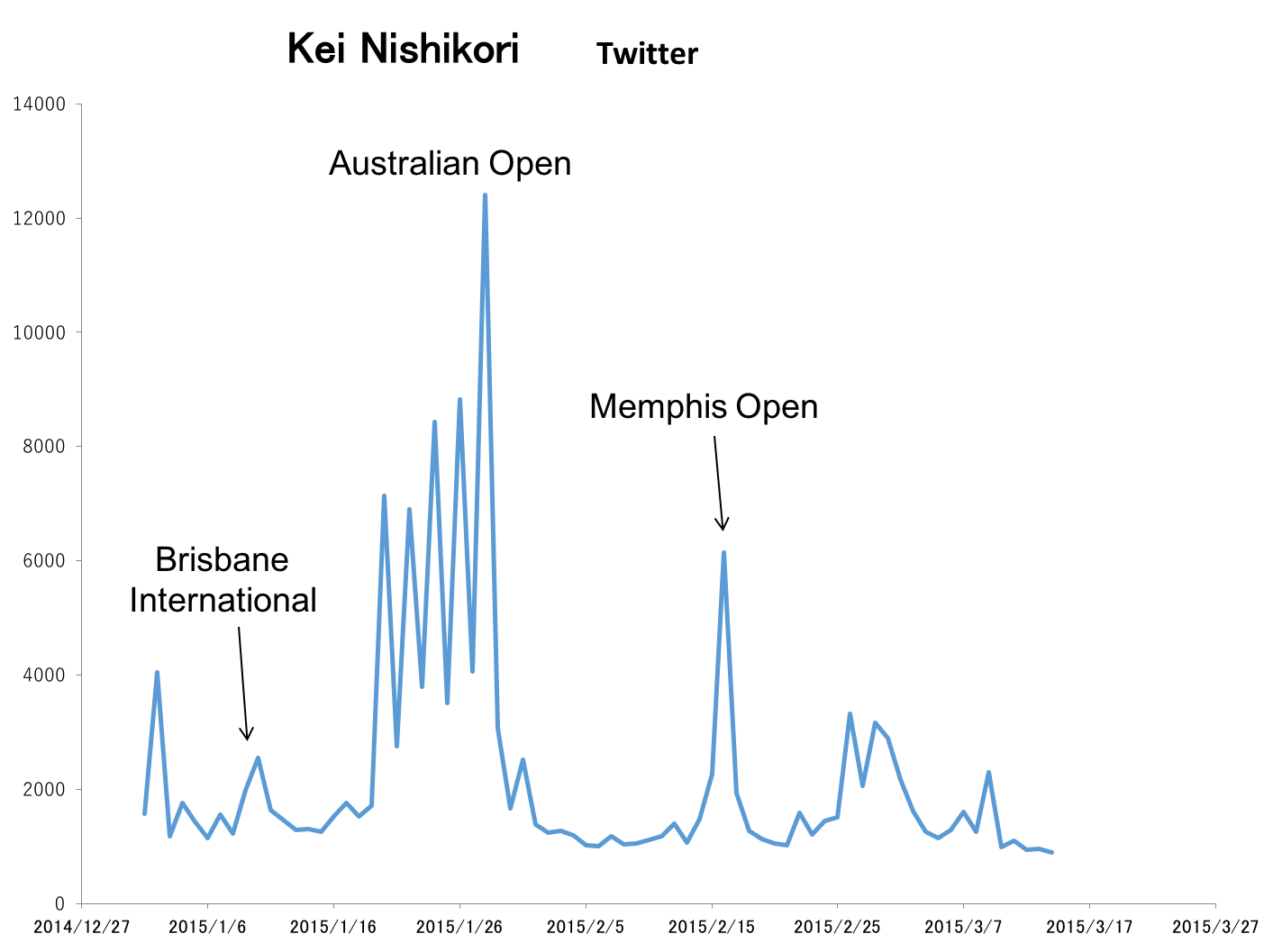}
\caption{Twitter of Nishikori Kei }
\label{fig:fig1}
\end{figure}

As a social physics theory to analyze society from the number of written social media, there is a mathematical model of the hit phenomenon. The mathematical model of the hit phenomenon is the theory of social physics submitted in 2012 by Ishii. \cite{Ishii2012a} There, we use equations of physics to present equations that explain people's interests from the influence of advertisements and the influence of communication with other people.

  In the mathematical theory for the hit phenomena, the effect of adverts and the propagation of reputation and rumors as a result of human communications are incorporated into the statistical physics of human dynamics. The mathematical model has been applied to the motion picture business in the Japanese market, and the calculations have been compared with the reported revenue and observed number of blog posts for each film. Furthermore, in  several recent papers, it was shown that the theory is not only applicable to the  box office, but also other social entertainment such as local events\cite{Ishii2012b}, animated dramas on TV\cite{Ishii2013b}, the ``general election'' of the Japanese girl-group AKB48\cite{Ishii2013a}, online music\cite{Ishii2012c}, plays\cite{Kawahata2013d}, music concerts\cite{Kawahata2013a,Kawahata2013b}, Japanese stage actors\cite{Kawahata2013e}, Kabuki players of the 19th century\cite{Kawahata2013c}, and TV dramas\cite{Ishii2014a}. In these works,  an extended mathematical theory for the hit phenomena was used to apply the model to general entertainment in society. Thus, it is very natural to use this theory for the prediction of motion picture business.
  
In this thesis, after showing a movie, drama, expansion of the topic of social incident etc using the mathematical model of hit phenomenon, we analyze the result of mathematical model of hit phenomena as analysis and prediction of social movement.　In this paper, responses in social media are observed using the social media listening platform presented by Hottolink. Using the dataset presented by M Data Co.\ Ltd., we can monitor the exposure of each film.

\section{Theory}

\subsection{mathematical model for the hit phenomenon}
 We write down the equation of purchase intention at the individual level  $I_i(t)$ as

 \begin{equation}
\frac{dI_i(t)}{dt} = \sum_{\xi} c_{\xi}A_{\xi}(t) - a I_i(t) + \sum_j d_{ij} I_j(t) + \sum_j \sum_k p_{ijk} I_j(t) I_k(t)
 \end{equation}

where $d_{ij}$, $h_{ijk}$, and $f_i(t)$ are the coefficient of the direct communication, the coefficient of the indirect communication, and the random effect for person i, respectively\cite{Ishii2012a}.  The advertisement and publicity effects are include in $A_{\xi}(t)$ which is treated as an external force. The index $\xi$ means sum up of the multi media exposures.  
Word-of-mouth (WOM) represented by posts on social network systems like blog or twitter is used as observed data which can be compared with the calculated results of the model. The unit of time is a day. 

Here, it is assumed that the height of interest I (t) of people attenuates exponentially. Although it is known that this is known to occur in movies and the like \cite{Ishii2012a}, attention such as events and anniversaries is known to attenuate by a power function. \cite{Sano2013a, Sano2013b} In the case of social interest, we attenuate the intermediate between the exponential function and the power function \cite{Ishii-Koyabu}, but here we simply adopt exponential decay.

We consider the above equation for every consumers in the society. Taking the effect of direct communication, indirect communication, and the decline of audience into account, we obtain the above equation for the mathematical model for the hit phenomenon. Using the mean field approximation, we obtain the following equation as equation for averaged intention in the society. The derivation of the equation is explained in detail  in ref.\cite{Ishii2012a}. 

\begin{equation}
\label{eq:eq13}
\frac{dI(t)}{dt} = \sum_{\xi} c_{\xi}A_{\xi}(t) + (D-a) I(t) + P I^2(t)
 \end{equation}

This equation is the macroscopic equation for the intention of whole society. Using this equation, our calculations for the Japanese motion picture market have agreed very well with the actual residue distribution in time \cite{Ishii2012a}. Using this equation, our calculations for the Japanese motion picture market agree very well with the actual residue distribution in time.  The advertisement and publicity effects are obtained from the dataset of M Data and the WOM represented by posts on social network systems are observed using the system of Hottolink.  We found that the indirect communication effect is very significant for huge hit movies. 

\subsection{Equilibrium solution}

Let the differential coefficient of I (t) be zero in equation (\ref{eq:eq13})  in order to obtain the equilibrium solution.

\begin{equation}
\label{eq:eq14}
 \sum_{\xi} c_{\xi}A_{\xi}(t) + (D-a) I(t) + P I^2(t) = 0
 \end{equation}

\begin{figure}[!h]
\centering
\includegraphics[height=6cm, bb=0 0 720 540]{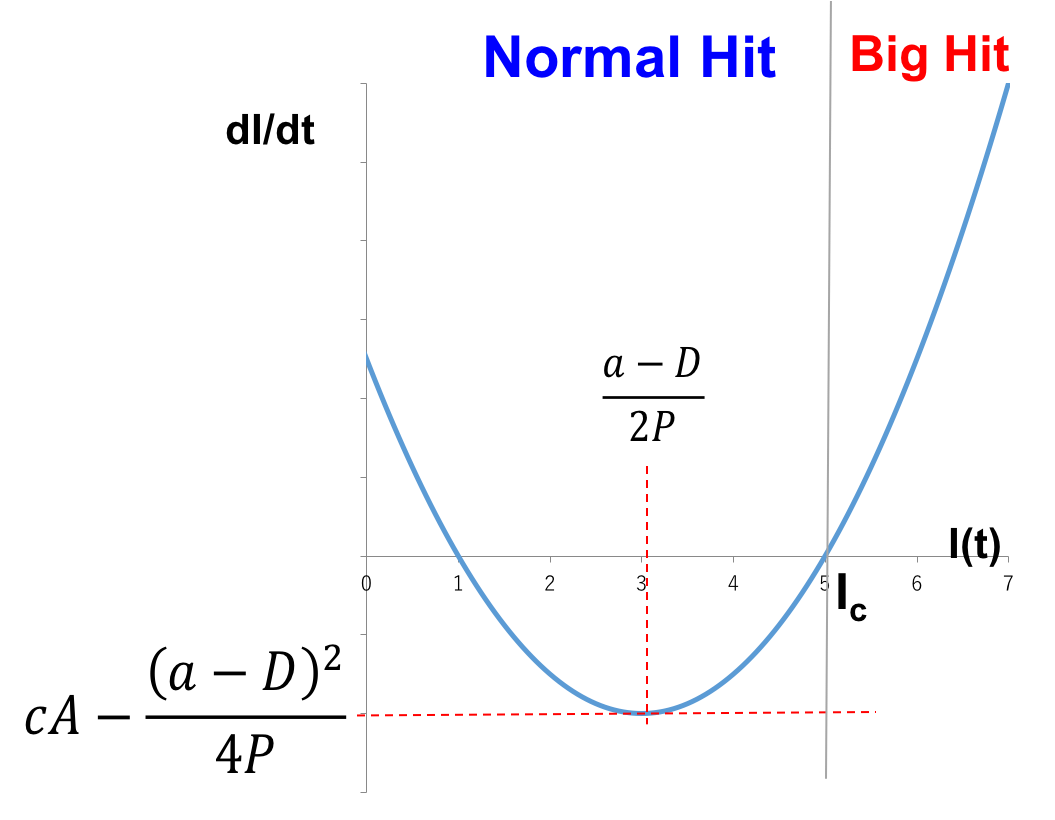}
\caption{Equilibrium solution of the equation (2). The region of "Big Hit" means the equilibrium solution diverges.  }
\label{fig:fig2}
\end{figure}

Let the differential coefficient of $I(t)$ be zero in equation (\ref{eq:eq13}) in order to obtain the equilibrium solution. There are two points of equilibrium. Let $I_c$ be the larger of the two. Graphs for $dI(t) / dt$ and $I(t)$ are shown in the figure \ref{fig:fig2}. From the figure, if $I(t) < I_c$, $I(t)$ converges to the point of the equilibrium solution, but if $I(t) > I_c$, $I(t)$ diverges. Actually, $I(t)$ does not diverge in real society from various constraints not included in the original equation, but it can be expected to be at least very large. This corresponds to a big hit phenomenon. In other words, if the interest in society is larger than a certain degree from the equation, it is included in the formula that it will become a big hit.

\subsection{Determination of parameters}

 For the purpose of reliability, we introduce the so-called ``R-factor'' (reliability factor), which is well known in the field of low-energy electron diffraction (LEED) \cite{Pendry}. In the LEED experiment, the experimentally observed curve of current vs.\ voltage is compared with the corresponding theoretical curve using the R-factor. 
 For our purpose, we define the R-factor as follows:

\begin{equation}
R = \frac{\sum_i (f(i)-g(i))^2}{\sum_i (f(i)^2 - g(i)^2)},
\end{equation}

where  $f(i)$ and $g(i)$ correspond to the calculated $I(t)$ and the observed number of blog posts  or tweets. The smaller the value of R, the better the functions $f$ and $g$. Thus, we use a random number to search for the parameter set that  minimizes R. This random number technique is similar to the Metropolis method\cite{Metropolis}, which we have used previously\cite{Ishii2012a}. We use this R-factor as a guide to obtain the best parameters for each calculation in this paper.

Actually, the parameters $c_{\xi}$, $D$, and $P$ in equation (\ref{eq:eq13}) can be considered as functions of time, because people's  attention changes over time. However, if  we introduce the functions $c_{\xi}(t)$, $D(t)$, and $P(t)$, we can tune any phenomena by adjusting these functions. Thus, we retain $c_{\xi}$, $D$, and $P$ as constant values to examine whether equation (\ref{eq:eq13}) can be applied to any social phenomena.

\section{Results}

Here, we will introduce an example of analyzing by applying mathematical model of hit phenomenon about television dramas, music concert, movies, and even social topic interests. And there are some cases where indirect communication is particularly large. The example analyzed in more detail will be shown at the end.

\subsection{Television dramas}

\begin{figure}[!h]
\centering
\includegraphics[height=6cm, bb=0 0 620 440]{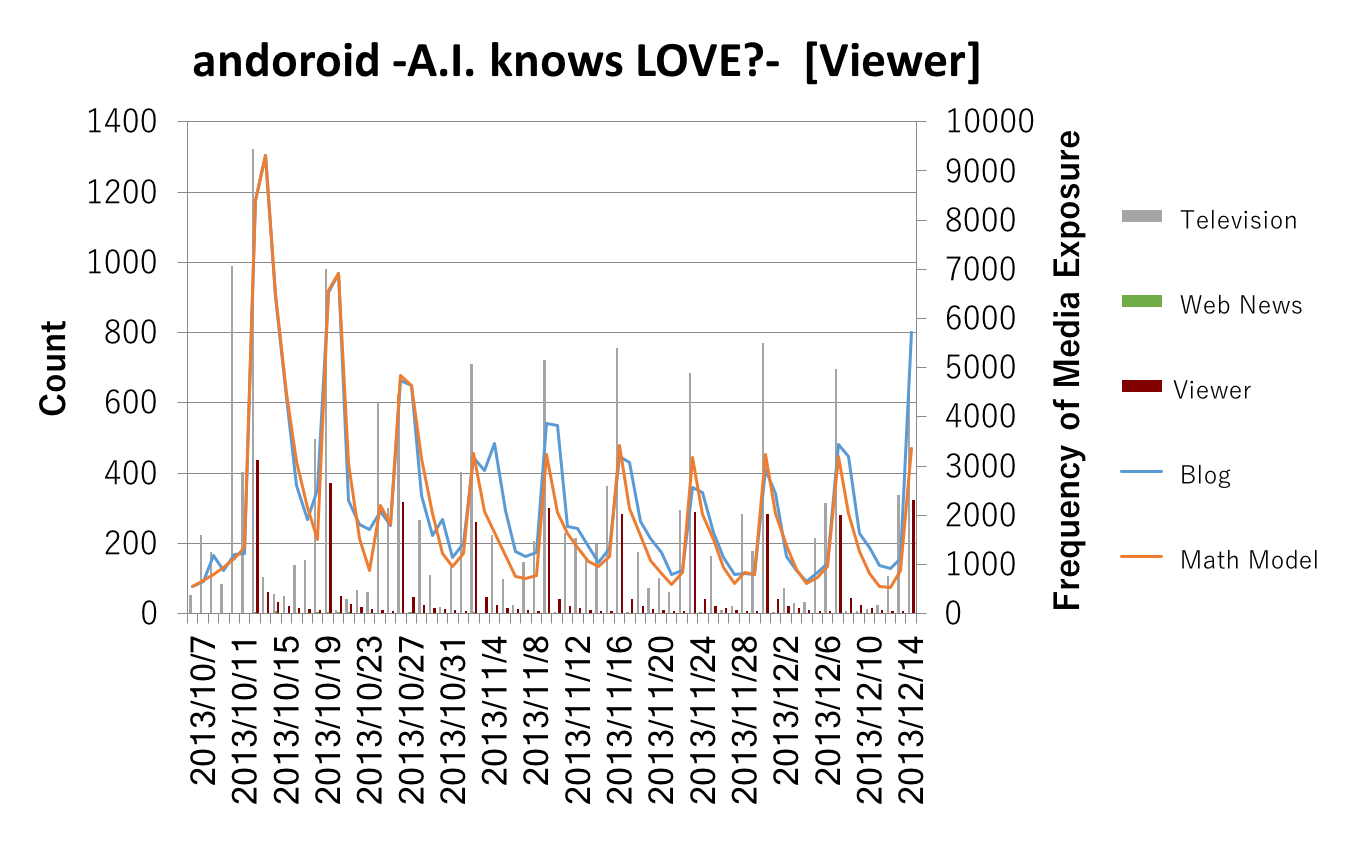}
\caption{The calculation and the observed data for  a weekly TV drama "android -A.I. knows LOVE?" in Japan. The histogram means the observed exposure on each movies on TV counted in unit of seconds.  The blue curve corresponds the observed number of daily blog posting and the red curve corresponds to our calculation. }
\label{fig:fig3}
\end{figure}

The first thing to mention is a TV drama\cite{Ishii2014a}. The figure \ref{fig:fig3} shows the analysis of Japan's "A.I. knows LOVE?" In 2013. The histogram in the figure is the number of seconds of exposure by advertisement or publicity on television, the majority of the exposure seconds is the drama broadcasting itself. As the number of social media writes, the number of blogs written was used. The peak corresponds to the broadcast of each time. The calculation result is calculated only in the first three times of all 10 drama shows, and the remaining seven times are prediction calculations. As can be seen, the prediction calculation has sufficient accuracy.

\subsection{Music concert}

\begin{figure}[!h]
\centering
\includegraphics[height=6cm, bb=0 0 720 350]{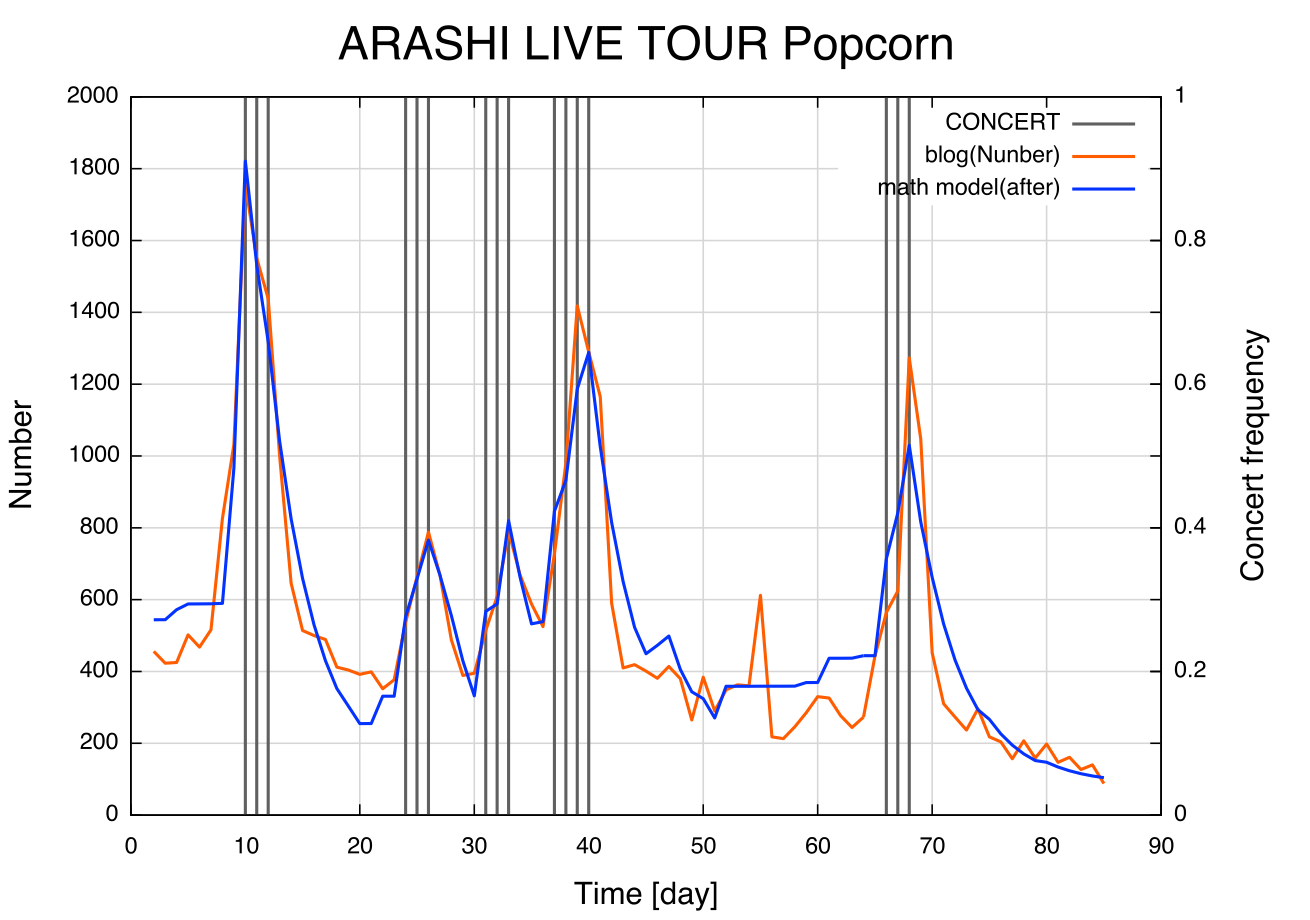}
\caption{ The calculation and the observed data for Music Concert of ARASHI  in Japan. The histogram means the observed exposure on each movies on TV counted in unit of seconds.  The red curve corresponds the observed number of daily blog posting and the blue curve corresponds to our calculation.}
\label{fig:fig4}
\end{figure}

In the figure \ref{fig:fig4}, the mathematical model of hit phenomenon is applied to music concerts. It was a very popular idol group called ARASHI in Japan. The peaks in the figure correspond to concerts by ARASHI respectively. As can be seen, the calculation results explain the number of social media written measured with sufficient precision, and the mathematical model of hit phenomenon can be applied to music concerts.

\subsection{Movie}

\begin{figure}[!h]
\centering
\includegraphics[height=6cm, bb=0 0 720 440]{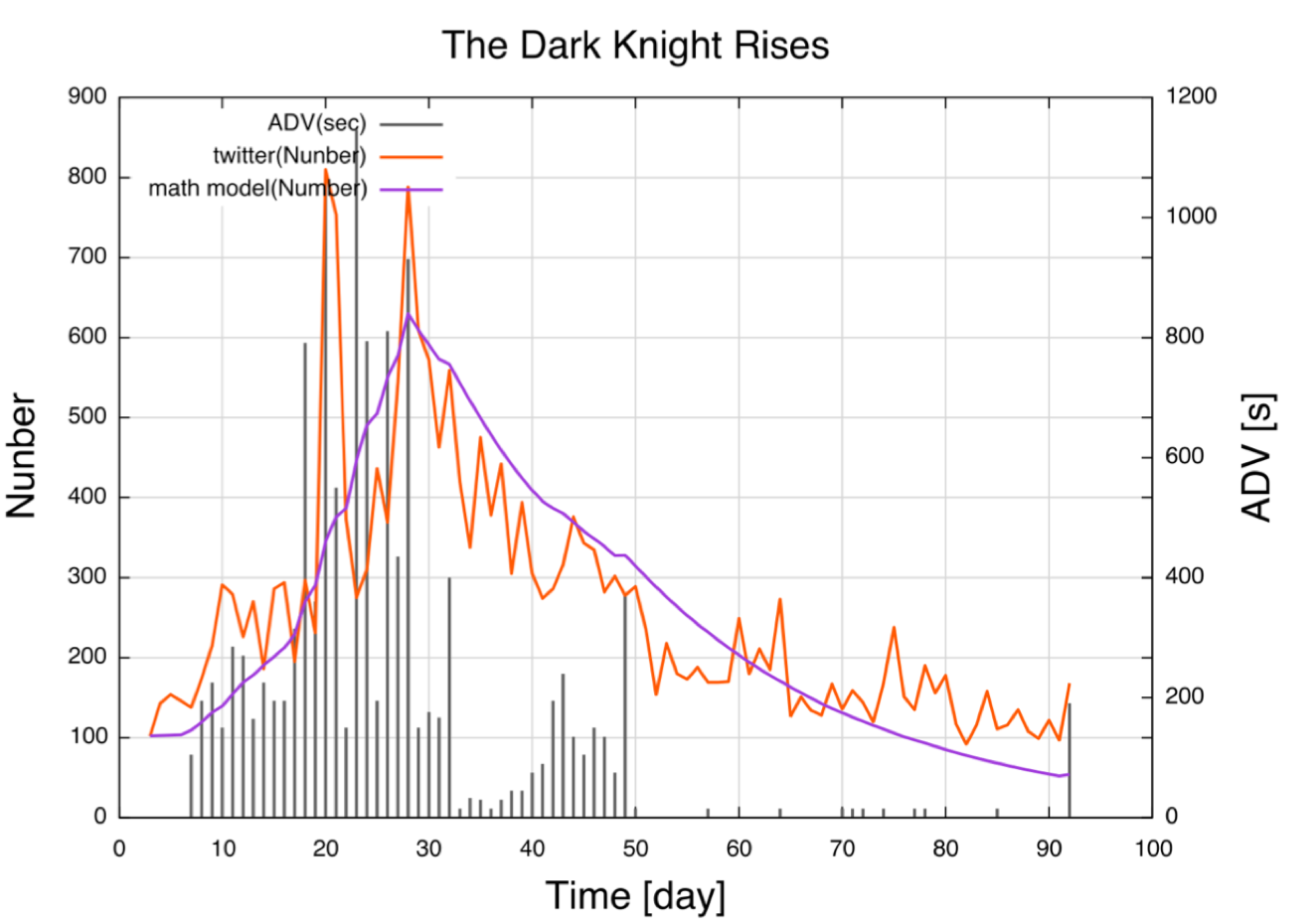}
\caption{ The calculation and the observed data for Movie "The Dark Knight Rises"  in Japan. The histogram means the observed exposure on each movies on TV counted in unit of seconds.  The orange curve corresponds the observed number of daily Twitter posting and the purple curve corresponds to our calculation. }
\label{fig:fig5}
\end{figure}

An example of a movie is shown in the figure \ref{fig:fig5}. Shown in the figure is the American work of 2012, "The dark knight rises". The histogram in the figure is the number of seconds of exposure by advertisement or publicity on television. Excluding small fluctuations, the calculation result explains Twitter's writing dorm accurately, and you can see that the mathematical model of the hit phenomenon is a theory that can be used for movies as well.

\subsection{Social scandals}

\begin{figure}[!h]
\centering
\includegraphics[height=6cm, bb=0 0 720 440]{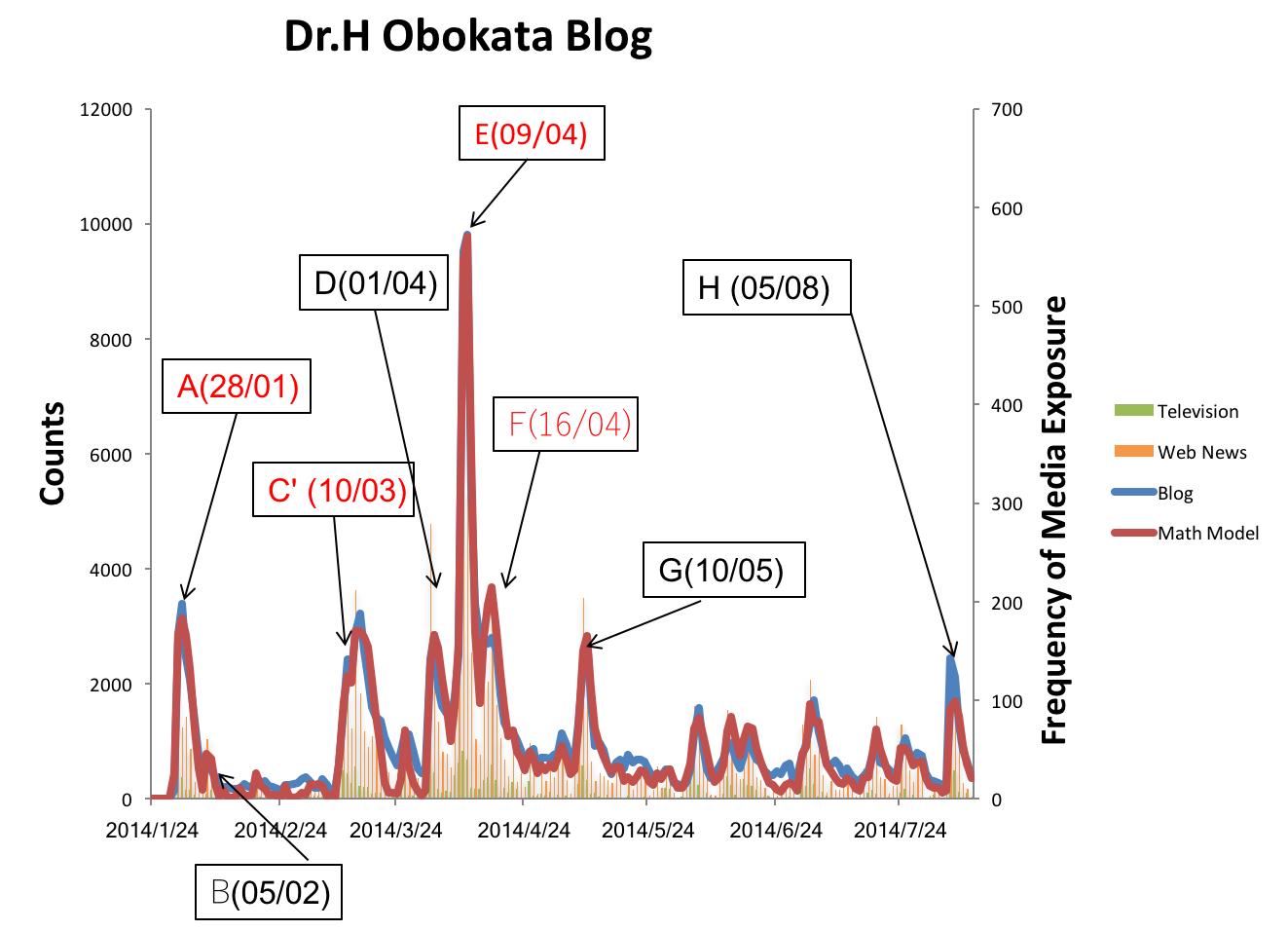}
\caption{Observation of daily posting to blog for "STAP cell" and corresponding calculation. The blue curve is the daily posting number and red curve is the calculation. The histograms are exposure on TV and internet news. A - H correspond to the description in text.  }
\label{fig:fig6}
\end{figure}

The next example is a scandal that made society troubled.\cite{Ishii2014b}  This was a case that happened in 2014 that it was actually foundation due to research fraud, which made a biological breakthrough called STAP cell, a great interest in Japanese society. The figure \ref{fig:fig6} shows the number of postings of daily postings of blogs written about Dr. Haruko Obokata who caused the STAP cell nebulization case and the calculation result by the mathematical model of the hit phenomenon describing it. In the time series in the figure, the days when major incidents occurred are as follows.

\begin{itemize}
\item A press conference of RIKEN for STAP cell (28 Jan)
\item B revelation of the suspicion (5 Feb)
\item C’ proposal of withdrawn of Nature paper by Prof.Wakayama(10 March)
\item C RIKEN released an interim report of the investigation on the suspicion (14　March)
\item D RIKEN released the final report for STAP cell suspicion (1 April)
\item E press conference by Dr. H Obokata (9 April)
\item F press conference by Prof. Y Sasai (16 April)
\item G Publication of the research notebooks of Obokata
\item H Decease of Prof. Y Sasai
\end{itemize}

As you can see from the figure the calculation shows quite good accuracy.

\begin{figure}[!h]
\centering
\includegraphics[height=6cm, bb=0 0 720 480]{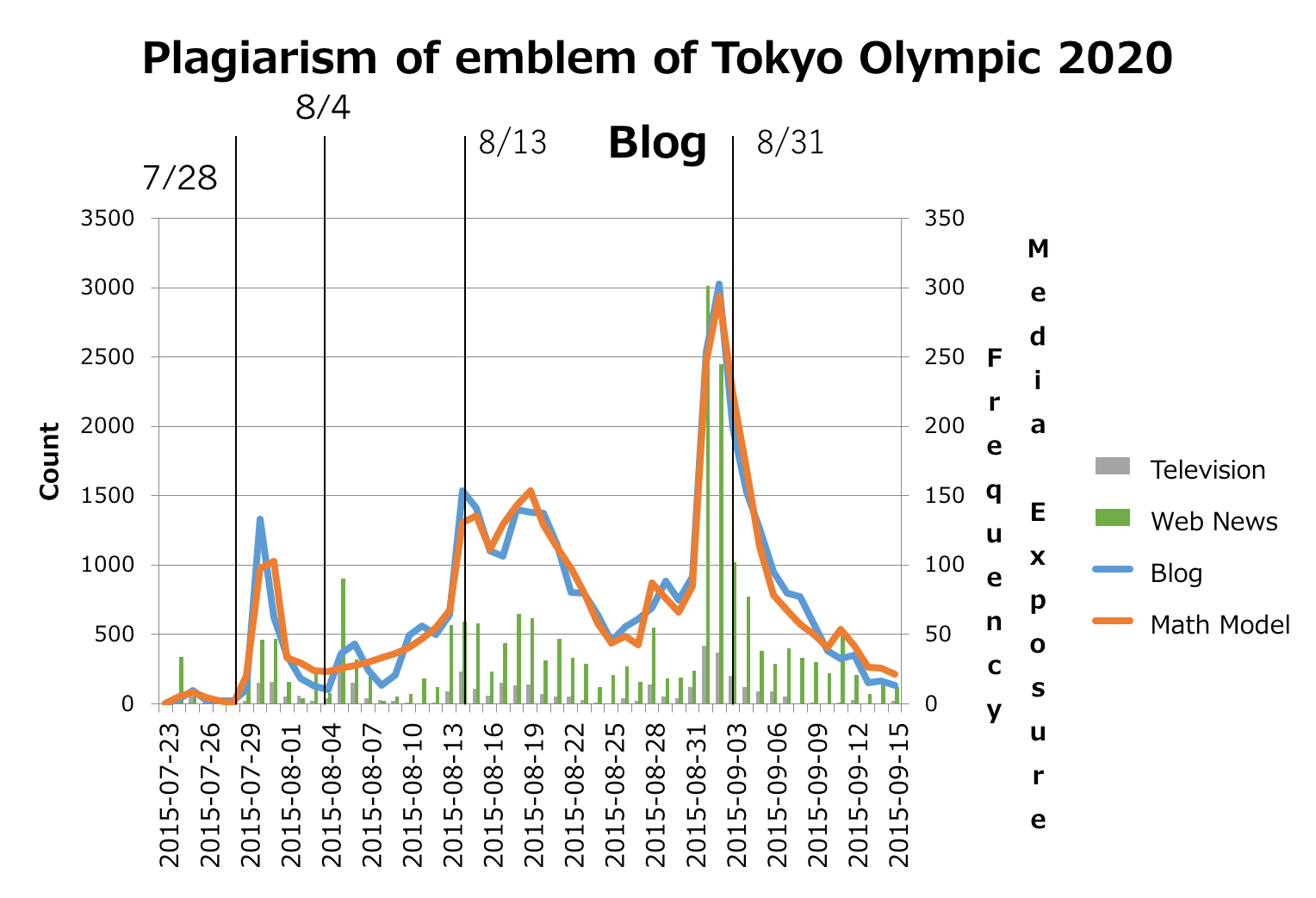}
\caption{ Observation of daily posting to blog for "Cancel of emblem of the Tokyo Olympic Games 2020" and corresponding calculation. The blue curve is the daily posting number and red curve is the calculation. The histograms are exposure on TV and internet news.  }
\label{fig:fig7}
\end{figure}

Another example was a case where the emblem of the Tokyo Olympic Games 2020 was canceled due to suspicion of design theft, happened in Japan in 2015. We measure the number of blog posts written about responsible designer Mr. Kenjiro Sano and compare it with the calculation result by the mathematical model of the hit phenomenon. As you can see in the figure \ref{fig:fig7}, the calculation results explain the number of blog posts with good accuracy.

\subsection{Strong indirect communication}

\begin{figure}[!h]
\centering
\includegraphics[height=6cm, bb=0 0 720 440]{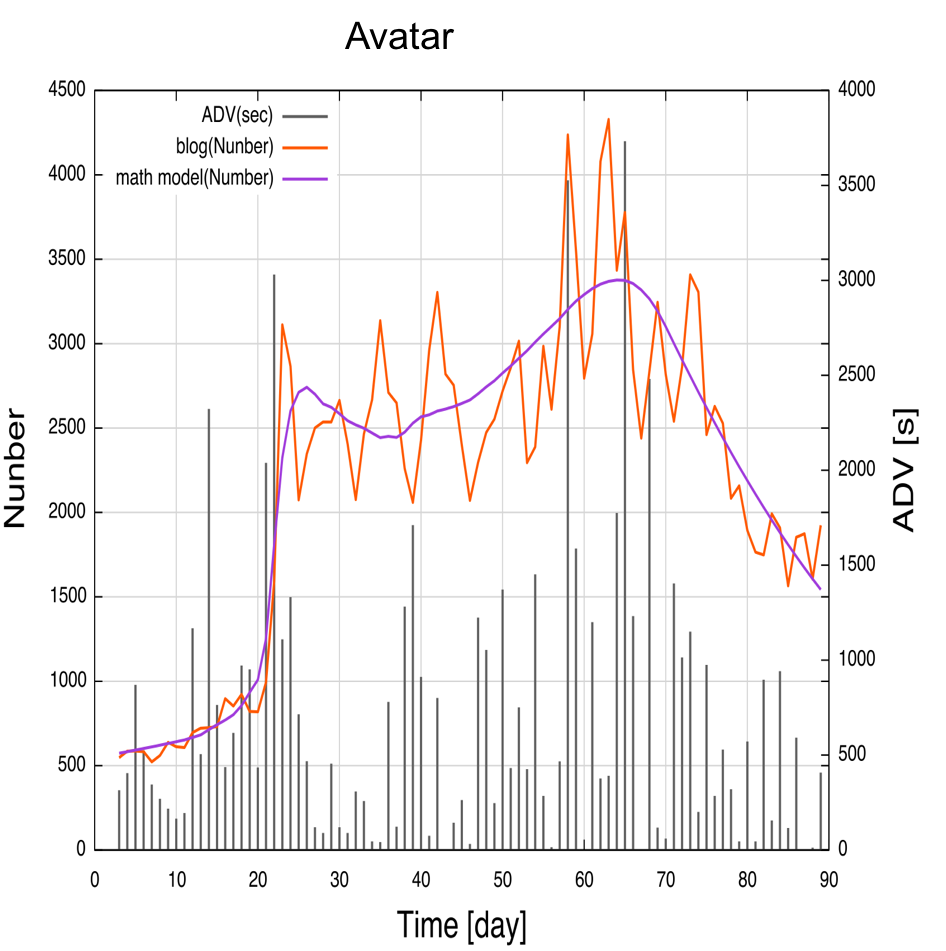}
\caption{Avatar }
\label{fig:fig8}
\end{figure}

\begin{figure}[!h]
\centering
\includegraphics[height=6cm, bb=0 0 720 440]{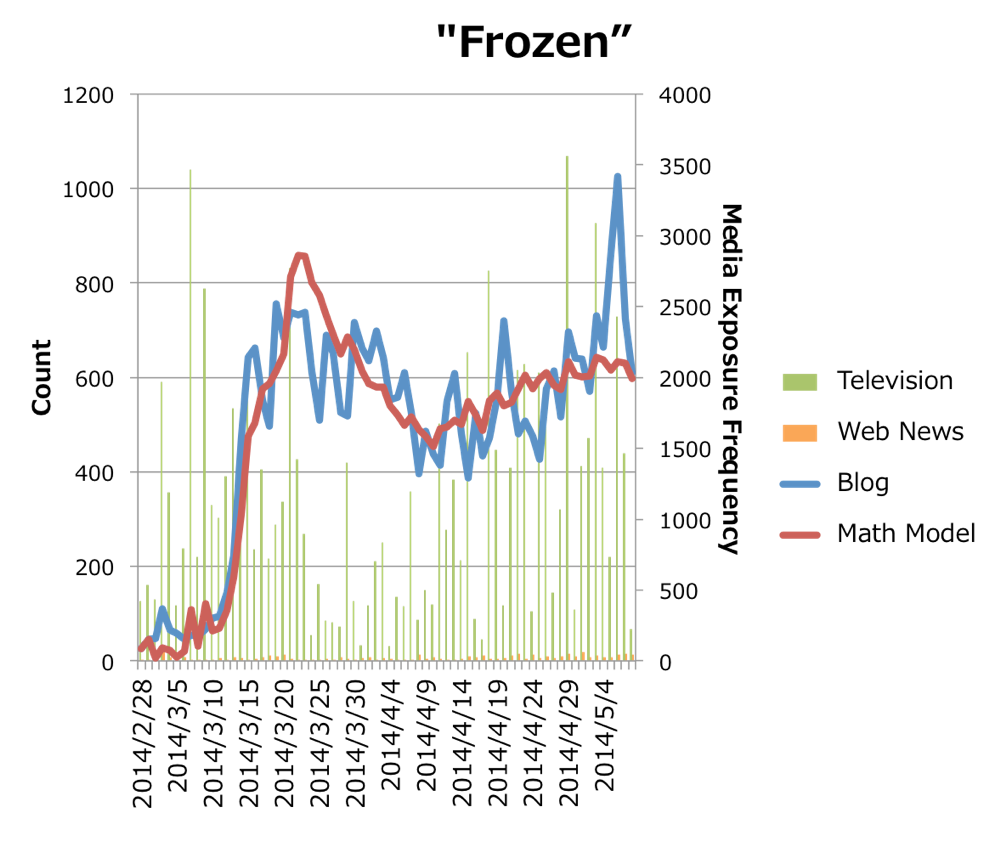}
\caption{Frozen}
\label{fig:fig9}
\end{figure}

In the analysis, let us show an example where indirect communication, which is a characteristic action in mathematical model of hit phenomenon, greatly affects. According to the mathematical model analysis of the hit phenomenon, movies show large indirect communication with several movies. Here are the Avatar (figure \ref{fig:fig8}) and Frozen (figure \ref{fig:fig9} ). The calculation result by the mathematical model of the hit phenomenon well matches the measurement of the blog post number. Large indirect communication is the cause of keeping movie hits long.

\subsection{Analysis using the mathematical model for hit phenomenon}

\begin{figure}[!h]
\centering
\includegraphics[height=6cm, bb=0 0 720 440]{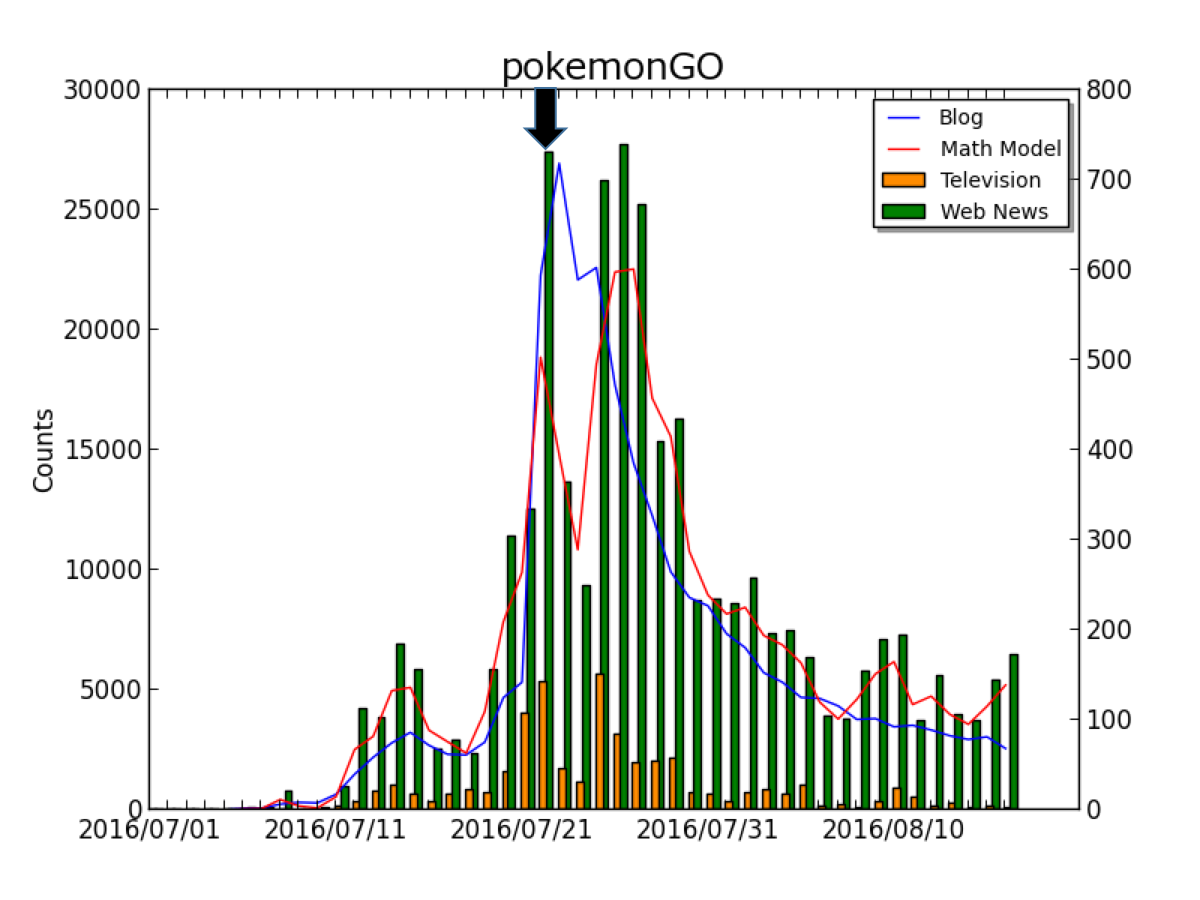}
\caption{Pokemon GO }
\label{fig:fig10}
\end{figure}

In the following example, the strength of direct communication and indirect communication is obtained from the calculation result by the mathematical model of the hit phenomenon, and it is used to analyze the movement of society, and the example is Pokemon GO.\cite{Ishii2016a} The figure \ref{fig:fig10} shows the number of blog posts on Pokemon GO about 1 month before and after July 22, 2016 when Pokemon GO began in Japan and the calculation result accordingly. Here we calculate separately before and after Pokemon GO starts in Japan. The results showed that direct communication became very strong after service started, while indirect communication became considerably weaker. The reason why direct communication became stronger is that the Pokemon GO game can actually be made, so it is probably that there are many conversations between users such as information exchange of pocket stops. The cause of indirect communication weakening may be thought as follows. Prior to the start of Pokemon GO, those who did not play games also had an interest in Pokemon GO 's enthusiasm in the US or in Europe. However, when games were actually available in Japan, those who did not play games would have lost their impression.

In this way, by calculating the mathematical model of hit phenomenon by delimiting time, it is possible to analyze whether direct communication or indirect communication becomes stronger or weakens.

\subsection{Prediction}

\begin{figure}[!h]
\centering
\includegraphics[height=6cm, bb=0 0 720 440]{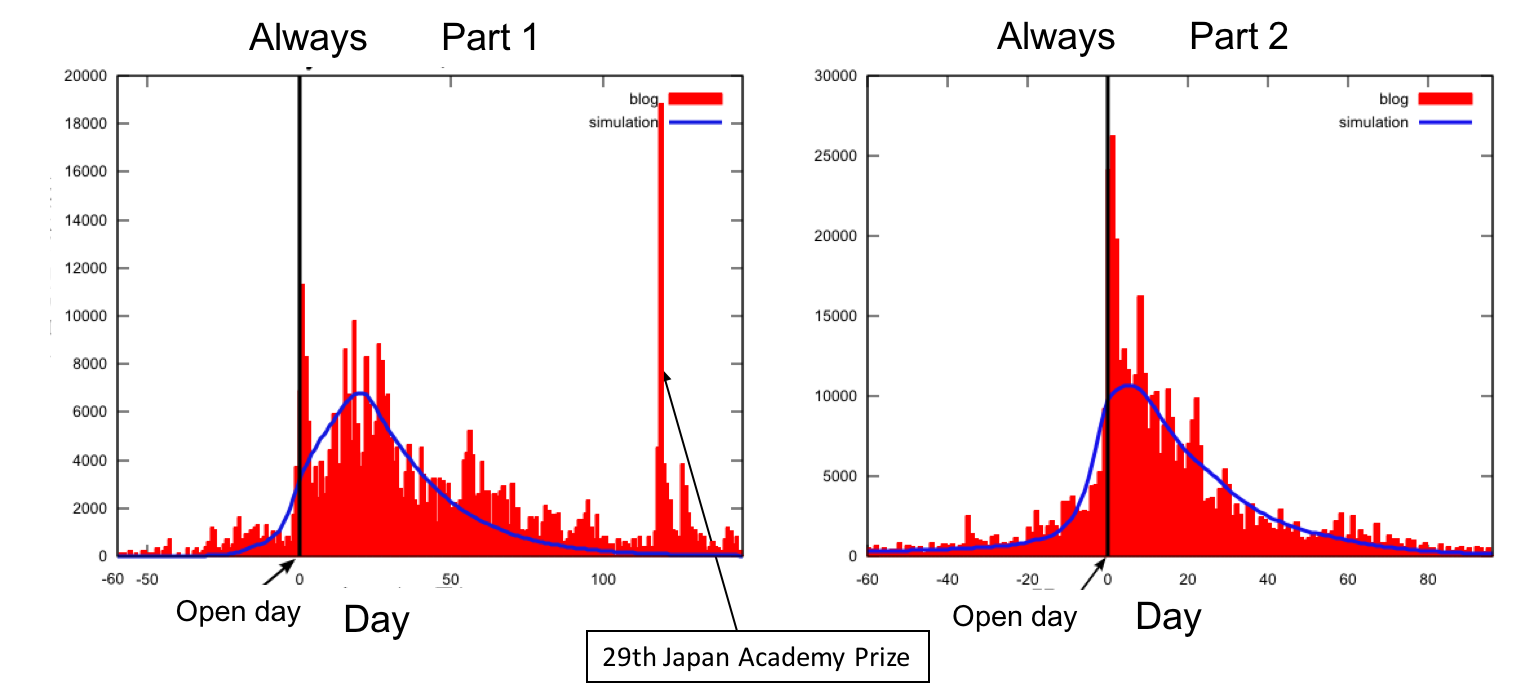}
\caption{Always Part 1 and Part 2}
\label{fig:fig11}
\end{figure}

Finally, let us show an example where prediction is possible by calculation using a mathematical model of hit phenomena, taking a movie as an example. Shown in the figure \ref{fig:fig11} is an example of calculating the reputation of Japanese movie called Always with a mathematical model of hit phenomenon. Calculate Part 1 in the same way as explained above, and determine parameters such as C, D, P from that. By using that parameter as it is to calculate Part 2, we were able to explain the number of blog posts with high precision by the mathematical model of the hit phenomenon with high accuracy without work such as changing the parameters and adapting the calculations in particular.

\begin{figure}[!h]
\centering
\includegraphics[height=6cm, bb=0 0 620 340]{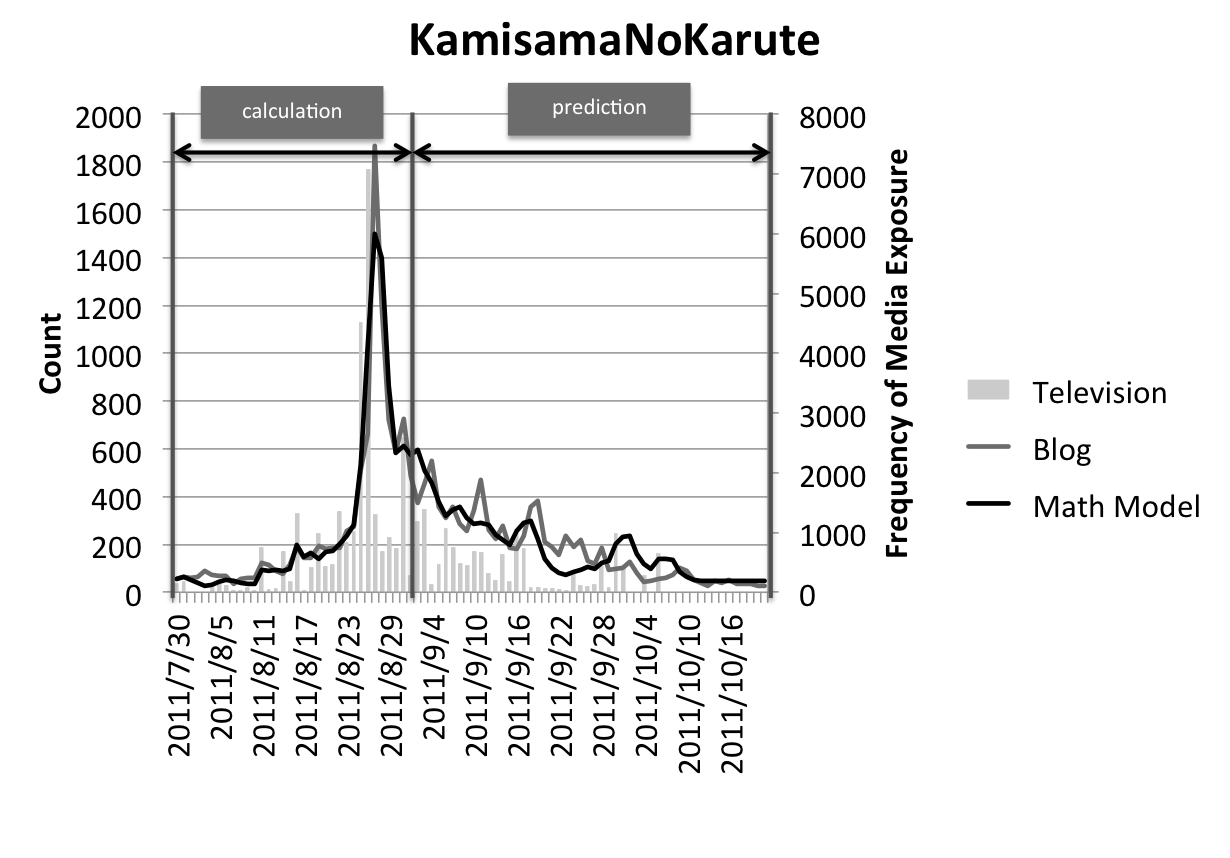}
\caption{Kamisama no Karte }
\label{fig:fig12}
\end{figure}

Also, a movie called "{\it Kamisama no Karte}" is calculated by a mathematical model of the hit phenomenon with data up to 4 days after release, and parameters, D, P are determined therefrom. It is figure \ref{fig:fig12} that it calculates after 5 days after release and matches it with the actual number of blog posts with the decided parameters. Although it is a forecast after the 5th day of release, it was able to predict with high accuracy.

\section{Discussion}

   The mathematical theory for the hit phenomena was presented, where the effect of adverts and the propagation of reputation and rumors as a result of human communications are incorporated into the statistical physics of human dynamics \cite{Ishii2012a}. The mathematical model for the hit phenomena' has been applied to the entertainment business in the Japanese market, and the calculations have been compared with the reported revenue and observed number of blog posts for each one. Moreover, in  several recent papers, as shown in the introduction section, we can apply the theory to many other social scandals. In these works,  an extended mathematical theory for the hit phenomena was used to apply the model to general entertainment in society. 

Using the mathematical model of the hit phenomenon, you can explain the transition of the number of blogs and Twitter 's written number by calculation results, as well as TV dramas, movies, music concerts, and scandals that made society troubles like STAP cell cases. The transition of the number of postings per day for any social events can be explained with high accuracy by theoretical calculation. The strength of direct communication and indirect communication obtained from calculation can be utilized for analysis of social movements. Direct communication is thought to be information transmission by conversation with the surroundings of each person and reflects the transmission of information within the cluster to which each person belongs.  On the other hand, from the original idea of indirect communication term in our mathematical model in ref.\cite{Ishii2012a}, indirect communication represents a case where information is transmitted beyond the cluster to which each person belongs. Therefore, indirect communication is strong in a big hit that will be talked about everywhere in society. On the other hand, as shown in Pokemon GO, when only some gamers speak hotly, direct communication alone is strong and indirect communication weakens.

Also, as shown in the movie example, prediction is also possible. In this way, the theory of social physics, which is a mathematical model of hit phenomena, can analyze reputation that becomes widely popular in society and predict attenuation and convergence. Therefore, it can be effectively used for marketing.

As a mathematical model of the hit phenomenon, as a theory of social physics, it is possible to describe how one person in the society will inflict interest and follow the time change of its interest. Therefore, expansion is easy. In order to deal with which of the two competing topics show interest, for example, a theory has been already proposed which makes two mathematical models of hit phenomena simultaneous.\cite{Ishii2016b, Ishii2016c}

It is also possible to solve the influence of social media on the market share of products by simultaneous theory of market share in economics and mathematical model of hit phenomenon.\cite{Ishii2016d}

\section{Conclusion}

In this paper, by mathematical model of hit phenomena, which is the theory of social physics, it is possible to calculate how topics will rise and converge in society, even to television dramas, movies, music concerts and social incidents we showed what we can do. Thus, we consider that mathematical model for hit phenomena is correct to explain spread of topics as social phenomena.   Using the mathematical model of hit phenomena, we can see whether there is spread of topics beyond clusters by social movements, and if indirect communication is big it will be a huge hit. Also, using the mathematical model of hit phenomena, prediction of reputation, that is, prediction of a big hit is also possible.

\subsection*{Acknowledgements}
The author would like to thank Dr.Yasuko Kawahata, Takuma Koyabu and 
Akiko Kitao for helpful suggestions on this paper.

\end{document}